\RequirePackage{ifpdf}
\ifpdf % We are running pdfTeX in pdf mode
\documentclass[pdftex]{sigma}
\else
\documentclass{sigma}
\fi

\numberwithin{equation}{section}

\begin{document}

\allowdisplaybreaks

\renewcommand{\thefootnote}{$\star$}

\renewcommand{\PaperNumber}{024}

\FirstPageHeading

\ShortArticleName{Generalized Heisenberg Algebras}

\ArticleName{Generalized Heisenberg Algebras, SUSYQM and\\ Degeneracies: Inf\/inite Well and Morse Potential\footnote{This
paper is a contribution to the Proceedings of the Workshop ``Supersymmetric Quantum Mechanics and Spectral Design'' (July 18--30, 2010, Benasque, Spain). The full collection
is available at
\href{http://www.emis.de/journals/SIGMA/SUSYQM2010.html}{http://www.emis.de/journals/SIGMA/SUSYQM2010.html}}}

\Author{V\'eronique HUSSIN~$^\dag$ and Ian MARQUETTE~$^\ddag$}

\AuthorNameForHeading{V.~Hussin and I.~Marquette}

\Address{$^\dag$~D\'epartement de math\'ematiques et de statistique,
Universit\'e de Montr\'eal,\\
\hphantom{$^\dag$}~Montr\'eal, Qu\'ebec H3C 3J7, Canada}
\EmailD{\href{mailto:veronique.hussin@umontreal.ca}{veronique.hussin@umontreal.ca}}

\Address{$^\ddag$~Department of Mathematics, University of York, Heslington, York YO10 5DD, UK}
\EmailD{\href{mailto:im553@york.ac.uk}{im553@york.ac.uk}}

\ArticleDates{Received December 23, 2010, in f\/inal form March 01, 2011;  Published online March 08, 2011}

\Abstract{We consider classical and quantum one and two-dimensional systems with ladder operators that satisfy generalized Heisenberg algebras. In the classical case, this construction is related to the existence of closed trajectories. In particular, we apply these results to the inf\/inite well and Morse potentials.  We discuss how the degeneracies of the permutation symmetry of quantum two-dimensional systems can be explained using  products of ladder operators.  These products satisfy interesting commutation relations. The two-dimensional  Morse quantum system is also related to a generalized two-dimensional Morse supersymmetric model. Arithmetical or accidental degeneracies of such system are shown to be associated to additional supersymmetry.}


\Classification{81R15; 81R12; 81R50}

\renewcommand{\thefootnote}{\arabic{footnote}}
\setcounter{footnote}{0}

\section{Introduction}

Considering one-dimensional  (1D) quantum \cite{Ma1} and classical \cite{Ma2} Hamiltonians with polynomial ladder operators (i.e.\ polynomial in the momenta) satisfying a polynomial Heisenberg algebra~\cite{FH, CFNN}, we have recently been able to construct two-dimensional (2D) superintegrable systems with separation of variables in cartesian coordinates, their integrals of motion and polynomial symmetry algebra. We have also discussed~\cite{Ma3} how supersymmetric quantum mechanics~\cite{Junker} can be used to generate new 2D superintegrable systems.

\looseness=-1
Superintegrable systems possess many properties that make them interesting from the point of view of physics and mathematics. Indeed, all bounded trajectories of maximally classical superintegrable systems are closed and the motion is periodic. Moreover, quantum 2D superintegrable systems have degenerate energy spectra explained by  Lie algebras,  inf\/inite dimensional algebras or polynomial algebras. For a review of 2D superintegrable systems we refer the reader to~\cite{Ma4}.

Many 1D quantum systems with nonlinear energy spectra such the inf\/inite well, the Scarf and the P\"oschl--Teller systems have ladder operators that satisfy a generalized Heisenberg algebra (GHA) \cite{Delange, DQ,EK1,EK2, GMK, Quesne, CZW, CR, DM, DK, Dong, CH, WYB}. These systems also appear in the context of 1D supersymmetric quantum mechanics (SUSYQM)~\cite{Junker}. It was also pointed out how some 2D generalizations of the Morse potential are related to 2D SUSYQM~\cite{IN}.

At the classical level, it was recently discussed how 1D systems like the Scarf, the inf\/inite well, the P\"oschl--Teller and the Morse systems allow such ladder operators that satisfy a Poisson algebra that is the classical analog of a GHA~\cite{KN}.  An interesting paper  \cite{CKN} compared the 1D classical and quantum P\"oschl--Teller potentials, their ladder operators and GHA. These ladder operators are directly related to time-dependent integrals of motion that give the motion in the phase space. In quantum mechanics ladder operators are well known and are used to generate, in particular, energy eigenstates, coherent and squeezed states \cite{Dong, DH,AH1, AH2}.

In this paper we are extending the constructions discussed in  \cite{Ma1, Ma2} to the inf\/inite well and Morse classical and quantum systems. Such systems are  integrable but not superintegrable. However some of their classical and quantum properties can be explained algebraically from their ladder operators as it is the case for superintegrable systems.

In Section~\ref{section2}, we recall the trajectories of the 1D classical inf\/inite well and Morse potentials and we give some examples of closed trajectories for 2D such systems. In Section~\ref{section3}, we present general ladder operators and GHA for the 1D classical systems and give them explicitly for our specif\/ic examples. We consider the extension to 2D of these systems and introduce some products of ladder operators which are not integrals of motion but satisfy interesting Poisson commutation relations. In fact, the commutation relations between these new operators can be interpreted as a condition to get closed trajectories in~2D. In Section~\ref{section4}, we recall the def\/initions of ladder operators and GHA for general 1D quantum systems. We consider again the examples of the inf\/inite well and  the Morse systems  \cite{Dong, DH, AH1, AH2, DLF} and we obtain the GHA. We also discuss, in Section~\ref{section5}, how using ladder operators, we can describe some degeneracies appearing from the permutation symmetry of isotropic 2D inf\/inite well and Morse potentials. The interpretation of accidental or arithmetical degeneracies from an additional supersymmetry is discussed in the case of the 2D Morse potential. We end the paper with some conclusions.

\section{Classical trajectories}\label{section2}

\subsection{Inf\/inite well system}\label{section2.1}

The trajectories for 1D and 2D inf\/inite well systems are given in details in~\cite{BMQ}. Here we just summarize the results to be able to use them later. In the 1D case, we take the Hamiltonian:
\begin{gather}\label{hamwell1}
H_{x}=\frac{P_{x}^{2}}{2m}+V(x),\qquad  V(x)= \begin{cases}
0 ,  & |x|<\dfrac{L}{2},\vspace{1mm} \\
 \infty , &   |x|>\dfrac{L}{2}.
\end{cases}
\end{gather}
The trajectories are given by
\begin{gather}\label{xt1}
x(t)=\begin{cases}
\dfrac{p_0}{m}t,  & t_{0} \leq t \leq t_{1}, \vspace{2mm}\\
 -\dfrac{p_0}{m}(t-t_{2\nu-1})+\dfrac{L}{2},  &  t_{2\nu-1}\leq t < t_{2\nu}, \vspace{2mm} \\
\dfrac{p_0}{m}(t-t_{2\nu})-\dfrac{L}{2},    &   t_{2\nu}\leq t < t_{2\nu+1},
\end{cases}
\end{gather}
where $n,\nu \in \mathbb{N}^{*}$ (i.e.\ are strictly positive integers) and $t_{n}=\frac{(2n-1)mL}{p_0}$. The initial conditions are chosen as $x_0=x(0)=0$ and $p_0=p(0)>0$ at $t=t_0=0$. The motion is periodic with period $T=\frac{4m L} {p_0}$ and frequency{\samepage
\begin{gather}
\omega=\frac{\pi p_0}{2mL}=\frac{\pi}{L}\sqrt{\frac{2E}{m}},\label{frequencywell}
\end{gather}
which is also written in terms of the total energy $E$ of the system under consideration.}

For the 2D system, we take, in particular, the Hamiltonian:
\begin{gather}
H=H_x+H_y=\frac{P_{x}^{2}}{2m}+\frac{P_{y}^{2}}{2m}+V(x)+V(y),\label{hamwell2}
\end{gather}
with
\begin{gather*}%\label{xt2}
V(x)= \begin{cases}
0 ,  & |x|<\dfrac{L}{2},\vspace{1mm} \\
 \infty , &   |x|>\dfrac{L}{2}
\end{cases},\qquad   V(y)= \begin{cases}
0 ,  & |y|<\dfrac{L}{2},\vspace{1mm}\\
 \infty , &   |y|>\dfrac{L}{2} .
\end{cases}
\end{gather*}
Trajectories are obtained directly from the 1D case and periodic orbits occur when $\omega_{x}n_{x}=\omega_{y}n_{y}$ where $n_{x}$ and $n_{y}\in \mathbb{N}^{*}$  \cite{BMQ}. Using the equation~\eqref{frequencywell} adapted to 2D, we obtain the following relation
\begin{gather}
n_{x}\sqrt{E_{x}}=n_{y}\sqrt{E_{y}}.\label{relenergywell}
\end{gather}
We show in Fig.~\ref{Fig1} a closed trajectory for the 2D inf\/inite well taking the initial position as  $(x_{0}, y_{0})=(0,0)$ and the initial momentum  as $p_0=(p_{0,x}, p_{0,y})$ with $p_{0,x}>0$, $p_{0,y}>0$. The parameters of the system are chosen as $m=\frac{L}{2}=1$.

\subsection{Morse system}\label{section2.2}

For the classical Morse system in 1D the trajectories have been obtained in~\cite{Slater}.  Let us recall some of the results. The Hamiltonian is given by
\begin{gather}
H_{x}=\frac{P_{x}^{2}}{2m}+V_{0}\big(e^{-2\beta x}-2 e^{-\beta x}\big).\label{hammorse1}
\end{gather}

The trajectories can be easily obtained from the Newton equation. For negative energy $E$ we get:
\begin{gather*}
x(t)=\frac{1}{\beta}\log\left(\frac{1-\cos\theta \cos(\omega t+\theta_{0})}{\sin^{2}\theta}\right),\qquad  \sin\theta=\sqrt{\frac{-E}{V_{0}}},%\label{xtmorse1}
\end{gather*}
where $\theta_{0}$ is a constant depending on the initial conditions. In the case of positive energy, the classical motion is given by a similar expression involving hyperbolic functions. Let us mention that the initial position~$x_{0}$ is dif\/ferent from zero while the initial momentum~$p_{0}$ is equal to zero. The motion for this 1D system is periodic and can be viewed as the logarithm of a harmonic oscillator. Indeed, the frequency at energy~$E$ is given by
\begin{gather}
\omega=\beta\sqrt{\frac{-2 E}{m}}.\label{frequencymorse}
\end{gather}

We can easily extend this resolution to the 2D system given, in particular, by the Hamiltonian:
\begin{gather}
H=H_x+H_y=\frac{P_{x}^{2}}{2m}+\frac{P_{y}^{2}}{2m}+V_{0}\big(e^{-2\beta x}-2 e^{-\beta x}\big)+V_{0}\big(e^{-2\beta y}-2 e^{-\beta y}\big).\label{hammorse2}
\end{gather}

\begin{figure}
 \begin{minipage}[b]{.46\linewidth}
 \centering
\includegraphics[width=65mm]{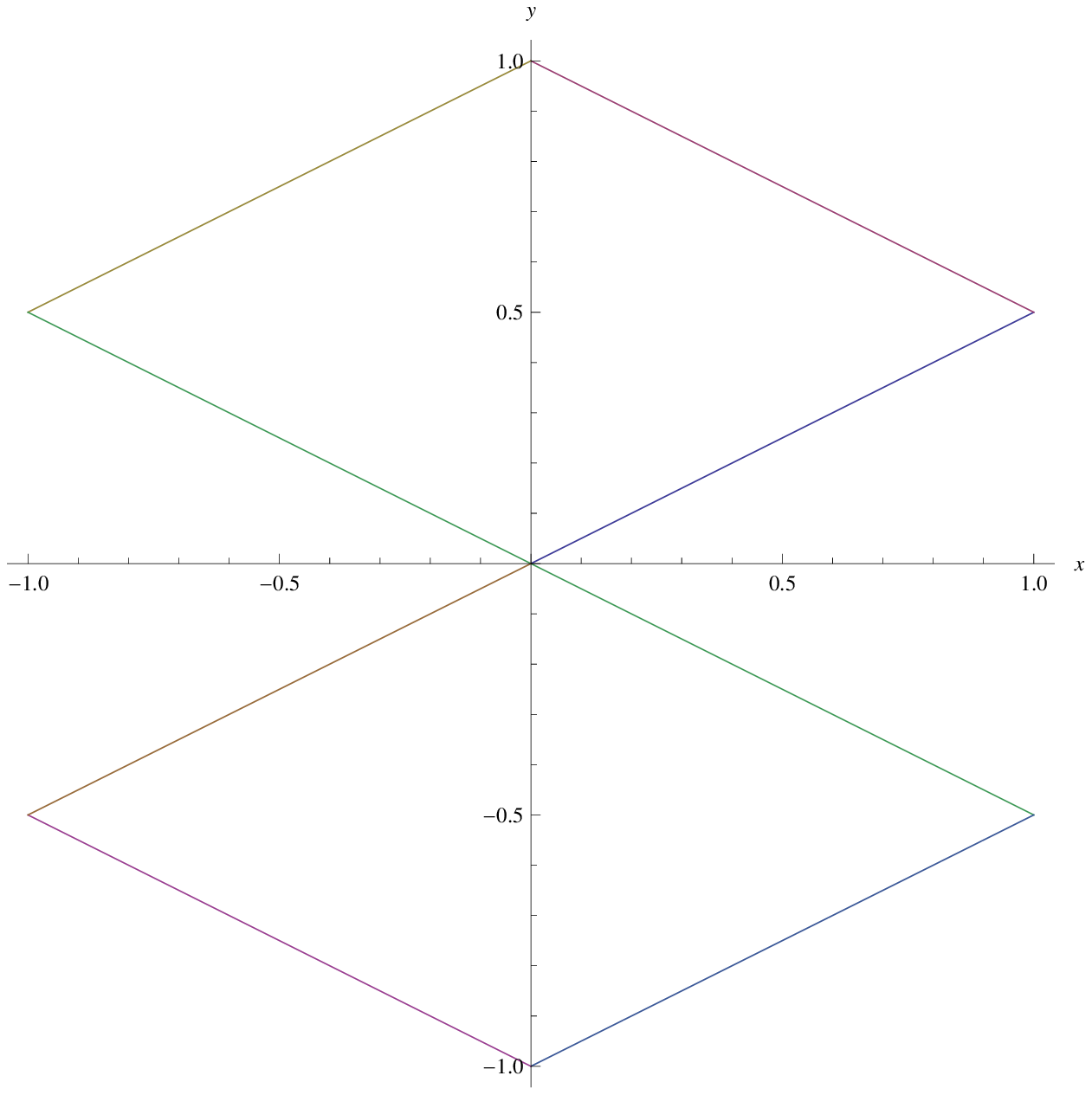}
% \centering\epsfig{figure=Fig1,width=\linewidth}
 \caption{A trajectory for $E_{x}=2$, $E_{y}=\frac{1}{2}$.}\label{Fig1}
 \end{minipage} \hfill
 \begin{minipage}[b]{.46\linewidth}
\centering
\includegraphics[width=65mm]{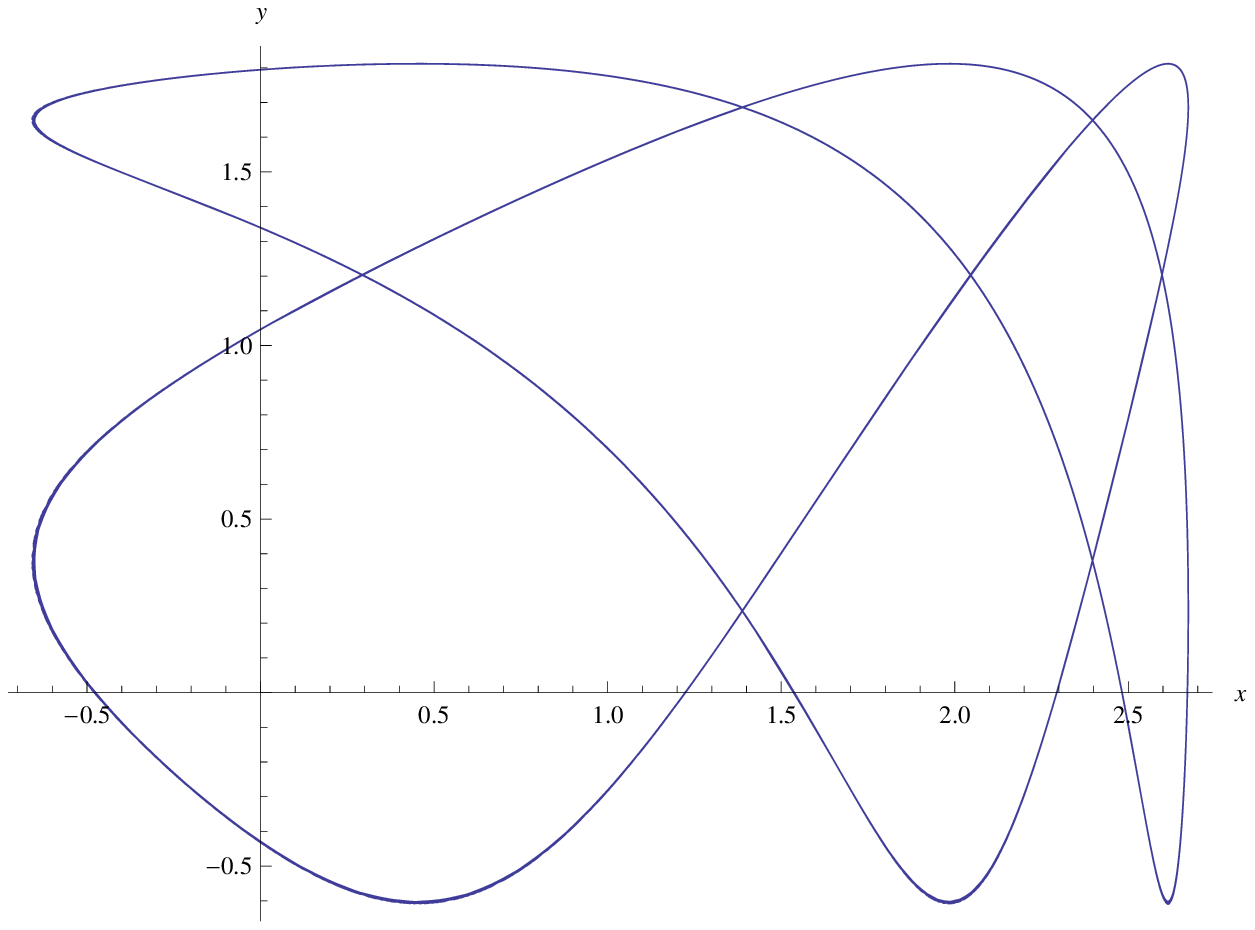}
% \centering\epsfig{figure=Fig2,width=\linewidth}
 \caption{A trajectory for $E_{x}=-2$, $E_{y}=-\frac{9}{2}$.}\label{Fig2}
 \end{minipage}
\end{figure}

As for the case of the inf\/inite well, closed trajectories and periodic orbits are obtained when $\omega_{x}n_{x}=\omega_{y}n_{y}$ or, from equation~\eqref{frequencymorse}, when the following condition
is satisf\/ied:
\begin{gather}
n_{x}\sqrt{-E_{x}}=n_{y}\sqrt{-E_{y}}, \qquad n_{x},  n_{y}\in \mathbb{N}^{*}.\label{relenergymorse}
\end{gather}
We obtain here the analog of Lissajous f\/igures for the anisotropic harmonic oscillator. One example is given in Fig.~\ref{Fig2} where we have chosen the energies~$E_{x}$ and~$E_{y}$ higher than the minimum of the potential and  such that equation~\eqref{relenergymorse} is satisf\/ied. The parameters of the system are taken as $m=\beta=1$ and $V_{0}=15$.

\section{Classical systems and generalized Heisenberg algebras}\label{section3}
We start by summarizing the factorization method in 1D classical mechanics  \cite{KN, CKN}. Let us assume that the system is given by the Hamiltonian $H$ and admits ladder operators of the following form
\begin{gather}
A^{\pm}=\pm i f(x)P_{x}+g(x) \sqrt{H}+\psi(x)+\phi(H),\label{laddergen}
\end{gather}
where the real functions $f$, $g$, $\psi$ are supposed to depend on~$x$ only and~$\phi$ is a real function of~$H$ (more precisely it is a power of~$\sqrt{H}$). Let us also assume that the set $\{H,   A^+,   A^-\}$ generates a~generalized Heisenberg algebra given as
\begin{gather}
\{H,A^{\pm}\}_{p}=\pm i \lambda(H)A^{\pm},\label{gha1}
\\
\{A^{+},A^{-}\}_{p}=-i \mu(H),\label{gha2}
\\
A^{+}A^{-}=H-\gamma(H),\label{gha3}
\end{gather}
where $\gamma(H)$, $\mu(H)$, $\lambda(H)$ are power of $\sqrt{H}$.  Note that the equation~\eqref{gha1} is a property that ladder operators must satisfy while the equation~\eqref{gha3} indicates a type of factorization of the Hamiltonian.
In the case of bounded motions with negative energy, we will replace the square root~$\sqrt{H}$ by~$\sqrt{-H}$.

Let us mention that additional Poisson commutation relations are satisf\/ied by the ladder operators of many classical systems \cite{KN, CKN}
\begin{gather}
\{\{H,A^{\pm}\}_{p},A^{\pm}\}_{p}=- 2a (A^{\pm})^{2}\label{gha5}
\end{gather}
or equivalently
\begin{gather}
\{\lambda(H),A^{\pm}\}_{p}=\pm 2ia A^{\pm}.\label{gha4}
\end{gather}

From equation~\eqref{gha1}, it is thus possible to construct  two time-dependent integrals of motion
\begin{gather}
Q^{\pm}=A^{\pm}e^{\mp i \lambda(H) t},\label{qpm}
\end{gather}
that indeed satisfy
\begin{gather*}
\frac{dQ^{\pm}}{dt}=\{H,Q^{\pm}\}_{p}+\frac{\partial Q^{\pm}}{\partial t}=0.%\label{constantqpm}
\end{gather*}
The frequency of the bounded states has thus an algebraic origin and is given by $\lambda(E)$. The constant values of these integrals $Q^{\pm}$ will be denoted by
\begin{gather}
q^{\pm}=c(E)e^{\pm i \theta_{0}}, \qquad c(E)=\sqrt{E-\gamma(E)}. \label{smallqpm}
\end{gather}
Indeed, the function $c(E)$ is obtained from the equation~\eqref{gha3} using the explicit form of the ladder operators given by the equation~\eqref{laddergen} and the form of $Q$ in equation~\eqref{qpm}
\begin{gather}
A^{\pm}=\pm i f(x)P_{x}+g(x)\sqrt{E}+\psi(x)+\phi(E)=c(E)e^{\pm i (\theta_{0}+\lambda(E) t)}.
\label{relladder}
\end{gather}
The trajectories in the phase space $(x(t),P_{x}(t))$ can thus be obtained algebraically from equation~\eqref{relladder}  with $c(E)=\eqref{smallqpm}$, as we will show in the examples below.

Now if we generalize the preceding considerations to a 2D system where the Hamiltonian allows the separation of variables in Cartesian coordinates
\begin{gather*}
H(x,y,P_{x},P_{y})=H_{x}+H_{y},%\label{hammorse2}
\end{gather*}
the system is integrable and possesses a second order integral of motion given by $S=H_{x}-H_{y}$.

Introducing the generalisation in 2D of the ladder operators~\eqref{laddergen} and denoting them~$A_{x}^{\pm}$ and~$A_{y}^{\pm}$, we can form the following products
\begin{gather*}
I_{+}^{(k,l)}=(A_{x}^{+})^{k}(A_{y}^{-})^{l},\qquad I_{-}^{(k,l)}=(A_{x}^{-})^{k}(A_{y}^{+})^{l},%\label{ipm}
\end{gather*}
where $k$ and $l \in \mathbb{N}^{*}$.
They satisfy
\begin{gather}
\big\{H,I_{\pm}^{(k,l)}\big\}_{p}=\pm i(k\lambda_{x}(H_{x})-l\lambda_{y}(H_{y}))I_{\pm}^{(k,l)}.\label{relipm}
\end{gather}
We have seen in \cite{FH} that for the special cases when  $\lambda_{x} (H_{x})$ and $\lambda_{y} (H_{y})$ are constant quantities and their ratio is rational,  the functions $I_{\pm}^{(k,l)}$ are thus integrals of motion. In such cases, systems are superintegrable and all bounded trajectories are closed. More generally, the condition for which the right side of equation~\eqref{relipm} vanishes corresponds to the condition for the existence of closed trajectories (see equation~\eqref{relenergywell}  or~\eqref{relenergymorse} in our particular examples)
\begin{gather*}
k\lambda_{x}((E_{x})-l\lambda_{y}((E_{y})=0.%\label{relenergywell2}
\end{gather*}

The 2D extension of the additional relations~\eqref{gha4}
implies the following constraints on $I_{\pm}^{(k,l)}$:
\begin{gather}
\big\{\big\{\big\{H,I_{\pm}^{(k,l)}\big\},I_{\pm}^{(k,l)}\big\},I_{\pm}^{(k,l)}\big\}_{p}=0.\label{relipm2}
\end{gather}

\subsection{Classical inf\/inite well}\label{section3.1}

The classical inf\/inite well may be included in the preceding algebraic scheme ($E>0$). Indeed, the ladder operators (with a slight modif\/ication) can be obtained from the quantum case  \cite{DM, Dong, CH} or as a limit of the classical P\"oschl--Teller system  \cite{KN, CKN}. The explicit form of the ladder operators for the inf\/inite well depends on the boundary conditions. Indeed, using the same boundary conditions as in Section~\ref{section2}, we get
\begin{gather}
A^{\pm}= \mp \frac{i}{\sqrt{2m}}\cos\left(\frac{\pi x}{L}\right)P_{x}+\sin\left(\frac{\pi x}{L}\right)\sqrt{H},\label{ladderwell}
\end{gather}
which generate the generalized Heisenberg algebra given in equations~\eqref{gha1},~\eqref{gha2} and~\eqref{gha3} with the following functions of $H$:
\begin{gather}
\lambda(H)= \mu(H)=\alpha \sqrt{H},\qquad \gamma(H)=0,\qquad \alpha =\frac{2\pi}{L\sqrt{2m}}.\label{functionswell}
\end{gather}
These ladder operators satisfy the equations~\eqref{gha4} (with $a=\frac{\alpha^{2}}{4}$).
The trajectories can thus be obtained algebraically from equations~\eqref{relladder} with $c(E)=\sqrt{E}$. Indeed the real part leads to
\begin{gather}
\sin\left(\frac{\pi x}{L}\right)=\cos\left(\theta_{0}+\frac{2\pi}{L\sqrt{2m}}\sqrt{E}  t\right)\label{relpositionwell}
\end{gather}
and thus, for the initial condition $x_{0}=0$ and the phase choice $\theta_{0}=-\frac{\pi}{2}$, we get
\begin{gather*}
x(t)=\sqrt{\frac{2E}{m}}t.%\label{xtwellfin}
\end{gather*}
We thus recover the equation~\eqref{xt1}. From the expression~\eqref{frequencywell} of the frequency, we indeed see that~$\lambda(E)$ is equal to~$\omega$.

Let us now consider the following 2D classical inf\/inite well as in equation (\ref{hamwell2}). We obtain from equations~\eqref{relipm} and~\eqref{functionswell}
\begin{gather}
\big\{H,I_{\pm}^{(k,l)}\big\}_{p}=\mp i \alpha \big(k \sqrt{H_{x}}-l\sqrt{H_{y}}\big)I_{\pm}^{(k,l)}.\label{relipmwell}
\end{gather}
These functions $I_{\pm}^{(k,l)}$ satisfy the equations~\eqref{relipm2}. We have thus shown that all  closed trajectories have an algebraic origin and the equations~\eqref{relipmwell} vanish when the equation~\eqref{relenergywell} is satisf\/ied.

\subsection{Classical Morse system}\label{section3.2}

The 1D Morse potential given by equation~\eqref{hammorse1} possesses the following classical ladder opera\-tors~\cite{KN} ($E<0$)
\begin{gather*}
A^{\pm}=\mp \frac{i}{\sqrt{2m}}e^{\beta x}P_{x}+e^{\beta x}\sqrt{-H}-\frac{V_{0}}{\sqrt{-H}}.%\label{laddermorse}
\end{gather*}
We thus get the generalized Heisenberg algebra with the following functions of $H$:
\begin{gather}
\lambda(H)= \epsilon \sqrt{-H},\qquad \mu(H)=\epsilon \frac{V_{0}^{2}}{H\sqrt{-H}},\qquad  \gamma(H)=V_{0}+H+\frac{V_{0}^{2}}{H},\qquad \epsilon = \frac{2\beta}{\sqrt{2m}}.\label{functionsmorse}
\end{gather}
These ladder operators satisfy the equations~\eqref{gha4} (with $a=-\frac{\epsilon^{2}}{4}$).

We also have ${c(E)=\sqrt{-\frac{V_{0}^{2}}{E}-V_{0}}}$. We thus get the trajectories algebraically and the frequency for the bound states are given by $\lambda(E)$ which is identical to equation~\eqref{frequencymorse}.

Let us now consider the 2D Morse potential given by equation~\eqref{hammorse2}. We obtain from equations~\eqref{relipm} and~\eqref{functionsmorse}
\begin{gather}
\{H,I_{\pm}^{(k,l)}\}_{p}=\pm i \epsilon (k \sqrt{-H_{x}}-l\sqrt{-H_{y}})I_{\pm}^{(k,l)}.\label{relipmmorse}
\end{gather}
The right side of the equation~\eqref{relipmmorse} vanishes again when equation~\eqref{relenergymorse} is satisf\/ied. All closed trajectories have an algebraic origin.

\section{Quantum systems with generalized Heisenberg algebras}\label{section4}

In 1D quantum mechanics a generalized Heisenberg algebra \cite{Delange, DQ,EK1,EK2, GMK, Quesne, CZW, CR, DM, DK, Dong, CH, WYB} generated by ladder operators $A^{\pm}$ and Hamiltonian~$H$ is def\/ined as the set $\{H, A^+,  A^-\}$ that satisf\/ies the commutation relations
\begin{gather}
[H,A^{+}]=A^{+}\lambda(H)=\delta(H)A^{+},\label{ghaq1}
\\
[H,A^{-}]=-\lambda(H) A^{-}=-A^{-}\delta(H)\label{ghaq2}
\end{gather}
and
\begin{gather}
[A^{+},A^{-}]=\gamma_{1}(H)-\gamma_{2}(H), \label{ghaq3}
\end{gather}
with
\begin{gather}
\quad A^{+}A^{-}=\gamma_{1}(H),\qquad A^{-}A^{+}=\gamma_{2}(H). \label{intergha3}
\end{gather}

The functions $\lambda(H)$,  $\delta(H)$, $\gamma_{1}(H)$, $\gamma_{2}(H)$ will take special expressions depending on the physical system under consideration.

It is well-known that, from equations~\eqref{ghaq1} or~\eqref{ghaq2}, we can obtain the energy spectrum algebraically. Indeed, let us consider the all set of eigenfunctions $\{|\psi_{n}\rangle, n\in I \}$ ($I$ is a f\/inite or inf\/inite set of positive integers) such that
\begin{gather}
H|\psi_{n}\rangle=E_{n}|\psi_{n}\rangle\label{eigengen1}
\end{gather}
and
\begin{gather}
A^{-}|\psi_{n}\rangle=\sqrt{k(n)}|\psi_{n-1}\rangle, \qquad  A^{+}|\psi_{n}\rangle=\sqrt{k(n+1)}|\psi_{n+1}\rangle.\label{ladderstates}
\end{gather}
From equations~\eqref{ghaq1},~\eqref{ghaq2} and~\eqref{ladderstates}, we get
\begin{gather*}
\lambda(E_{n})=\delta(E_{n+1})=E_{n+1}-E_{n}%\label{lambda}
\end{gather*}
and, from equations~\eqref{intergha3} and~\eqref{ladderstates},
\begin{gather*}
\gamma_{1}(E_{n})=k(n), \qquad \gamma_{2}(E_{n})=k(n+1).%\label{gamma}
\end{gather*}
For the physical systems we are considering, the energy spectrum is at most quadratic. This implies that if we take
\begin{gather}
E_n=a n^2+b n+c, \label{energyquad}
\end{gather}
we easily get
\begin{gather}
\lambda(E_{n})=\delta(E_{n})+2a=a(2n+1)+b.\label{lambdadelta}
\end{gather}
Moreover, it is possible to show that,
\begin{gather*}
[[H,A^{\pm}],A^{\pm}]=2a (A^{\pm})^2%\label{doublecom}
\end{gather*}
or, equivalently,
\begin{gather}
[\lambda(H),A^{\pm}]=\pm 2a A^{\pm},\label{lambdacom}
\end{gather}
which is the quantum analog of the constraint (\ref{gha5}) or (\ref{gha4}).

Let us mention that if we take
\begin{gather*}
A_{0}=-\frac{1}{2a} \lambda(H),%\label{azero}
\end{gather*}
and assume that
\begin{gather}
\gamma_{1}(H)- \gamma_{2}(H)=\frac{1}{a} \lambda(H),\label{gamma}
\end{gather}
the ladder operators $A^\pm$, together with the new generator $A_{0}$, generate a $su(1,1)$ algebra
\begin{gather*}
[A^{\pm}, A_{0}]=\pm A^{\pm}, \qquad [A^{+},A^{-}]=-2 A_{0}. %\label{com11}
\end{gather*}

When we act on the eigenfunctions $|\psi_{n}\rangle$, we also get
\begin{gather}
[H,(A^{\pm})^k]=(E_{n\pm k}-E_{n}) (A^{\pm})^k.\label{ladderk}
\end{gather}

Now we consider again 2D systems as a sum of 1D systems that have ladder operators generating the preceding algebraic structure. The Hamiltonian is written as:
\begin{gather}
H(x,y,P_{x},P_{y})=H_{x}(x,P_{x})+H_{y}(y,P_{y}).\label{hamq2}
\end{gather}
It admits a set of eigenfunctions labelled as $\{|\psi_{n_{x},n_{y}}>, n_{x}, n_{y} \in I \}$  such that
\begin{gather*}
H|\psi_{n_{x}, n_{y}}\rangle =E_{n_{x}, n_{y}}|\psi_{n_{x}, n_{y}}\rangle \qquad {\rm{with}} \quad E_{n_{x}, n_{y}}= E_{n_{x}}+  E_{n_{y}}.%\label{eigengen}
\end{gather*}

As in the classical case, the separation of variables allows the existence of a second order integral of motion $S=H_{x}-H_{y}$. We can again take the products of the ladder operators as ($k, l\in \mathbb{N}^{*}$)
\begin{gather}
I_{+}^{(k,l)}=(A_{x}^{+})^{k}(A_{y}^{-})^{l},\qquad I_{-}^{(k,l)}=(A_{x}^{-})^{k}(A_{y}^{+})^{l}.\label{ipmq}
\end{gather}

Let us impose the following constraints on the ladder operators (as in the 1D case with equation~\eqref{lambdacom})
\begin{gather*}
[\lambda(H_{x}),A_{x}^{\pm}]=\pm 2a A_{x}^{\pm},\qquad [\lambda(H_{y}),A_{y}^{\pm}]=\pm 2a A_{y}^{\pm}. %\label{lambdaq1}
\end{gather*}

From  equations~\eqref{ladderk} and~\eqref{ipmq}, we get
\begin{gather}
[H,I_{\pm}^{(k,l)}]=\pm I_{\pm}^{(k,l)}(k\lambda_{x}(H_{x})-l\lambda_{y}(H_{y})+k(k-1)a+l(l+1)a).   \label{comipm}
\end{gather}

Indeed, using equation (\ref{ladderk}), we get, for example,
\begin{gather*}
[H,I_{+}^{(k,l)}]|\psi_{n_{x},n_{y}}\rangle=((E_{n_{x}+k}-E_{n_{x}})+(E_{n_{y}-l}- E_{n_{y}})) I_{+}^{(k,l)} |\psi_{n_{x},n_{y}}\rangle.   %\label{comipm1}
\end{gather*}
Due to the fact that the total energy $E_n$ takes the general form  (\ref{energyquad}) and $\lambda(E_{n})$ is thus given by equation (\ref{lambdadelta}), we easily get
\begin{gather*}
(E_{n_{x}+k}-E_{n_{x}})+(E_{n_{y}-l}- E_{n_{y}})= k (2 a n_{x}+b)-l (2 a n_{y}+b)+a \big(k^2+l^2\big)\nonumber \\
\hphantom{(E_{n_{x}+k}-E_{n_{x}})+(E_{n_{y}-l}- E_{n_{y}})}{}
 = k(\lambda_{x}(E_{n_{x}})-a)-l(\lambda_{y}(E_{n_{y}})-a)+a \big(k^2+l^2\big). %\label{commHI}
\end{gather*}
This is the quantum equivalent of the equation~\eqref{relipm} for systems with quadratic energy spectrum. Similarly to the classical case, when $\lambda_{x}(H_{x})$ and $\lambda_{y}(H_{y})$ reduce to constants and their ratio is rational, the operators $I_{\pm}^{(k,l)}$ are thus integrals of motion and the Hamiltonian given by equation~\eqref{hamq2} is superintegrable~\cite{Ma1, Ma2}.

For our specif\/ic case of a quadratic spectrum, we have seen that the expression of $\lambda$ is given by equation~\eqref{lambdadelta} and the commutators  $[H,I_{^\pm}^{(k,l)}]$ are zero if $a(2kn_{x}-2ln_{y}+k^{2}+l^{2})+b(k-l)=0$. In particular, for $k=l$ and $a\neq 0$, we get $n_{y}=n_{x}+k$.

We will discuss, in the next section how, when $\lambda_{x}(H_{x})$ and $\lambda_{y}(H_{y})$ do not reduce to a constant, we can explain some of the degeneracies of the energy spectrum.

Let us f\/inally mention that we also obtain from equation~\eqref{comipm} the following commutation relations
\begin{gather*}
\big[\big[\big[H,I_{\pm}^{(k,l)}\big],I_{\pm}^{(k,l)}\big],I_{\pm}^{(k,l)}\big]=0,%\label{tricom}
\end{gather*}
which are the quantum analogs of relations given by equations~\eqref{relipm2}.

\subsection{Inf\/inite well potential}\label{section4.1}

We consider the quantum version of the 1D inf\/inite well given in equation~\eqref{hamwell1}. The Schr\"odinger equation and the corresponding energy spectrum are well-known and given by (using $\alpha$ as def\/ined in equation~\eqref{functionswell})
\begin{gather*}
H|\psi_{n}\rangle=E_{n}|\psi_{n}\rangle,\qquad E_{n}=\frac{\alpha^{2} \hbar^{2}}{4}n^{2} , \qquad n=1,2,\dots.%\label{eigenwell1}
\end{gather*}
To simplify the following expression, we introduce the number operator $N$ that satisf\/ies $N|\psi_{n}\rangle=n |\psi_{n}\rangle$ and is thus def\/ined in terms of $H$ as
\begin{gather*}
N= \frac{2}{\alpha\hbar} \sqrt{H}.%\label{nofh}
\end{gather*}
It has been shown that the ladder operators $A^\pm$ may be def\/ined as \cite{DM,Dong,CH}
\begin{gather*}
A^{-} = \sqrt{1+\frac{2}{N}}\left(-\frac{iL}{\pi}P_{x}\cos\left(\frac{\pi x}{L}\right)-N\sin\left(\frac{\pi x}{L}\right)\right), \\ %\label{amoinswell}\\
A^{+} = \left(\cos\left(\frac{\pi x}{L}\right)\frac{iL}{\pi}P_{x}-\sin\left(\frac{\pi x}{L}\right)N\right)\sqrt{1+\frac{2}{N}}. %\label{apluswell}
\end{gather*}
Let us mention the similarity with respect to the classical case and the equation~\eqref{ladderwell}. These ladder operators satisfy the relations~\eqref{ladderstates} with
\begin{gather*}
k(n)=n^{2}-1.%\label{knwell}
\end{gather*}

We can thus form a generalized Heisenberg algebra given by \eqref{ghaq1},  \eqref{ghaq2} and  \eqref{ghaq3} with
\begin{gather*}
\lambda(H) = \alpha\hbar\sqrt{H}+\frac{\alpha^{2}\hbar^{2}}{4}=\frac{\alpha^{2}\hbar^{2}}{4}(2N+1),\\ %\label{functionswellq} \\
  \gamma_{1}(H) = \frac{4}{\alpha^{2}\hbar^{2}}(H-E_{1}),\qquad \gamma_{2}(H)=\gamma_{1}-\frac{4}{\alpha^{2}\hbar^{2}}\lambda(H).%\label{functionswellq1}
\end{gather*}

Note that the operator $\lambda(H)$ satisf\/ies the relations of commutation given by equation~\eqref{lambdacom} with $a=\frac{\alpha^{2}\hbar^{2}}{4}$.
We also see that the relation~\eqref{gamma} is satisf\/ied giving rise to the well-known existence of a $su(1,1)$ algebra associated to the inf\/inite well system.

\subsection{Morse potential}\label{section4.2}

We consider the 1D quantum version of the Hamiltonian~\eqref{hammorse1}. Introducing the  parameters
\begin{gather*}
\nu=\sqrt{\frac{8mV_{0}}{\beta^{2}\hbar^{2}}}, \qquad p=\frac{\nu-1}{2},%\label{parameter}
\end{gather*}
we get the Schr\"odinger equation and the corresponding energy spectrum
\begin{gather}
H|\psi_{n}\rangle=E_{n}|\psi_{n}\rangle,\qquad E_{n}=\frac{-\hbar^{2}\beta^{2}}{2m}\left(\frac{\nu-1}{2}-n\right)^{2}=-\frac{\hbar^{2}\epsilon^{2}}{4}(p-n)^{2},\label{eigenmorse}
\end{gather}
where $n=0,1,\dots,[p]$. Indeed, the last admissible value of~$n$ is the integer part of~$p$. Again
the number operator $N$ satisf\/ies $N|\psi_{n}\rangle=n |\psi_{n}\rangle$ and  is def\/ined in terms of~$H$ as
\begin{gather*}
N=p-\frac{2}{\hbar\epsilon}\sqrt{-H}.%\label{nofhmorse}
\end{gather*}

We introduce the following ladder operators  \cite{DH, AH1}
\begin{gather*}
A^{-}=\left(\frac{e^{\beta x}}{\left(p+\frac{1}{2}\right)\beta}\left(\frac{i}{\hbar}P_{x}+\frac{\beta}{2}(2p -2N)(2p - 2N+1)\right)+\left(p+\frac{1}{2}\right)\right) \sqrt{K(N)},%\label{amoinsmorse}
\\
A^{+}=\big(\sqrt{K(N)}\big)^{-1}\left(\frac{e^{\beta x}}{\left(p+\frac{1}{2}\right)\beta}\left(-\frac{i}{\hbar}P_{x}+\frac{\beta}{2}(2p -2N)(2p - 2N-1)\right)-\left(p+\frac{1}{2}\right)\right),%\label{aplusmorse}
\end{gather*}
where
\begin{gather*}
K(N)=\frac{(2p-N)(2p-2N+2)}{(2p-N+1)(2p-2N)}.%\label{Kn}
\end{gather*}
They act on the eigenfunctions of the Morse potential as in equation~\eqref{ladderstates} where
\begin{gather*}
k(n)=n(2p-n).
%\label{eq63}
\end{gather*}
We can again form a generalized Heisenberg algebra given by equations~\eqref{gha1},  \eqref{gha2} and  \eqref{gha3} with
\begin{gather*}
\lambda(H) = \hbar\epsilon\sqrt{-H}-\frac{\hbar^{2}\epsilon^{2}}{4}=\frac{\hbar^{2}\epsilon^{2}}{4}(2p-2N-1),\\ %\label{functionsmorseq} \\
\gamma_{1}(H) = \frac{4}{\hbar^{2}\epsilon^{2}}(H-E_{0}), \qquad \gamma_{2}(H)=\gamma_{1}(H)+\frac{4}{\hbar^{2}\epsilon^{2}}\lambda(H)%\label{functionsmorseq1}
\end{gather*}
and the operator $\lambda(H)$ satisf\/ies the equation~\eqref{lambdacom} with $a=\frac{-\hbar^{2}\epsilon^{2}}{4}$. Here again, we can construct a corresponding $su(1,1)$ algebra.

\section{Degeneracies of quantum systems\\ with generalized Heisenberg algebras}\label{section5}

The study of the relation between the degeneracies and symmetries of multi-dimensional systems consisting in a sum of 1D systems with nonlinear energy spectrum~\cite{Itzykson} appears to be a~less studied subject than in the case of linear energy spectrum. This is a consequence of the fact that multi-dimensional systems that are sum of 1D systems with linear spectrum possess the superintegrability property~\cite{Ma1, Ma2, Ma3, Ma4} and appear in the classif\/ication of superintegrable systems. The most well-known of such systems is the anisotropic harmonic oscillator.

Only the 2D anisotropic $Q$-oscillators was discussed in~\cite{WYB} and the permutation symmetry was obtained algebraically using operators $\{I_{+},I_{-},I_{3}\}$ (def\/ined as series of ladder operators) generating a $su(2)$ algebra and commuting with the Hamiltonian. This symmetry manifests itself only in doublet and singlet states.

We will see how, from results of Section~\ref{section4.2}, such type of operators may be considered as well for the cases of the isotropic 2D inf\/inite well and Morse potentials.

These last systems present also accidental or arithmetic degeneracies~\cite{Itzykson} that has not been explained algebraically. Some of the arithmetical degeneracies of the 2D inf\/inite well were presented in~\cite{DH} and references therein. We are proposing an interpretation for the case of the Morse potential based on the existence of supercharges of second order. In fact, this example is particularly interesting because this interpretation will also be valid for the super partner of the 2D Morse potential which is not separable in cartesian coordinates (not even in any either coordinates).

For the 2D quantum systems we are considering, we have seen that they admit a quadratic energy spectrum of the form
\begin{gather*}
E_{n_{x},n_{y}}=E_{n_{x}}+E_{n_{y}}=a\big(n_{x}^2+n_{y}^2\big)+b(n_{x}+n_{y})+2c.
\end{gather*}
In particular, for the isotropic inf\/inite well, we have $a=\frac{\hbar^{2}\alpha^{2}}{4}$ and $b=c=0$ while for the Morse system, we have $a=c=\frac{-\hbar^{2}\epsilon^{2}}{4}$ and $b=\frac{\hbar^{2}\epsilon^{2}p}{2}$. The energies of these systems have similar algebraic structure and present the two types of degeneracies mentioned before. The f\/irst type appears when we make the change $n_{x} \leftrightarrow n_{y}$. It is associated to the permutation symmetry. The second type corresponds to the fact that two or more dif\/ferent sets $\{(n_{i,x},n_{i,y}), i\in \mathbb{N}^{*}\}$ could give rise to the same energy. It is called accidental or arithmetic degeneracy~\cite{Itzykson}.

\subsection{Permutation degeneracies}\label{section5.1}

Let us consider the Fock space and the action of the ladder operators as given in equations~\eqref{eigengen1} and~\eqref{ladderstates} generalized for 2D. We use here the simplif\/ied notation $|n_{x},n_{y}\rangle$ for the general energy eigenstates $|\psi_{n_{x},n_{y}}\rangle$. We thus denote the operators $I_{\pm}^{(k,l)}$ def\/ined in~(\ref{ipmq}) by  $I_{\pm}^{(l)}$ when $k=l$. We see that the operator  $I_{+}^{(l)}$ takes a state with $n_{x}=i$ and $n_{y}=i+l$ to a state with $n_{x}=i+l$ and $n_{y}=i$ and the operator $I_{-}^{(l)}$ takes a state with $n_{x}=i+l$ and $n_{y}=i$ to a state with $n_{x}=i$ and $n_{y}=i+l$.

For the inf\/inite well, we thus consider  the following operators:
\begin{gather*}
I_{+} = \sum_{i=0}^{\infty}\sum_{l=1}^{\infty}\frac{k(i)!}{k(i+l)!}\big(I_{+}^{(l)}\big)|i,i+l\rangle\langle i,i+l| =\sum_{i=0}^{\infty}\sum_{l=1}^{\infty}|i+l,i\rangle\langle i,i+l|,\\  %\label{Ipluswell}\\
I_{-}=I_{+}^{\dagger} = \sum_{i=0}^{\infty}\sum_{l=1}^{\infty}\frac{k(i)!}{k(i+l)!}|i,i+l\rangle\langle i,i+l|\big(I_{-}^{(l)}\big) =\sum_{i=0}^{\infty}\sum_{l=1}^{\infty}|i,i+l\rangle\langle i+l,i|,%\label{Imoinswell}
\end{gather*}
and
\begin{gather*}
I_{3}=\frac{1}{2}\sum_{i=0}^{\infty}\sum_{l=1}^{\infty}(|i+l,i\rangle\langle i+l,i|-|i,i+l\rangle\langle i,i+l|).%\label{I3well}
\end{gather*}
They commute with the Hamiltonian of the system and describe the permutation symmetry in the energy spectrum of the 2D isotropic inf\/inite well potential that manifests itself only in doublets and singlets. Indeed, we get
\begin{gather*}
I_{+}|j,k\rangle=|k,j\rangle, \qquad  I_{-}|j,k\rangle=|k,j\rangle,\qquad I_{3}|j,k\rangle=0.
\end{gather*}

For the Morse potential, the number of bound states is f\/inite, we thus take only a f\/inite sum for such kind of operators. We get
\begin{gather*}
I_{+} = \sum_{i=0}^{[p]}\sum_{l=1}^{[p]-i}\frac{k(i)!}{k(i+l)!}\big(I_{+}^{(l)}\big)|i,i+l\rangle\langle i,i+l|=\sum_{i=0}^{[p]}\sum_{l=1}^{[p]-i}|i+l,i\rangle\langle i,i+l|,\\ %\label{Iplusmorse}\\
I_{-}=I_{+}^{\dagger} = \sum_{i=0}^{[p]}\sum_{l=1}^{[p]-i}\frac{k(i)!}{k(i+l)!}|i,i+l\rangle\langle i,i+l|\big(I_{-}^{(l)}\big) =\sum_{i=0}^{[p]}\sum_{l=1}^{[p]-i}|i,i+l\rangle\langle i+l,i|,%\label{Imoinsmorse}
\end{gather*}
and
\begin{gather*}
I_{3}=\frac{1}{2}\sum_{i=0}^{[p]}\sum_{l=1}^{[p]-i}(|i+l,i\rangle\langle i+l,i|-|i,i+l\rangle\langle i,i+l|).%\label{I3morse}
\end{gather*}
These products of ladder operators can thus be related to symmetries of the system
as it is the case for superintegrable systems.

\subsection{Arithmetical degeneracies}\label{section5.2}

We are considering the example of the 2D Morse system in order to give an interpretation of arithmetical degeneracies in terms of symmetry or more precisely supersymmetry of the system.
Indeed, the existence of a supersymmetric partner of the original system has been shown by Iof\/fe and collaborators (see~\cite{IN} and references therein for more details and generalizations). We will thus start this subsection by introducing such a partner and the corresponding supercharges which realize the intertwining~\cite{IN}.

From equation~\eqref{eigenmorse} generalized for 2D, we get the energy spectrum of the Hamiltonian $H=H_{x}+H_{y}$ as
\begin{gather}
E_{n_{x},n_{y}}=E_{n_{x}}+E_{n_{y}}=-\frac{\hbar^{2}\epsilon^{2}}{4}\big((p-n_{x})^{2}+(p-n_{y})^2\big), \label{energy12}
\end{gather}
with $n_{x},n_{y}=0,1,2,\dots,[p]$. The super partner of $H$ has been obtained as
\begin{gather*}
\tilde H=H+\frac{\hbar^{2}}{m}\frac{\beta^2}{2 \sinh^2\left(\frac{\beta (x-y)}{2}\right)}.
\end{gather*}
Indeed, introducing the supercharges $Q^\pm$, which are dif\/ferential operators of order~2 in the momenta,
\begin{gather*}
Q^\pm =-H_{x}+H_{y}+D^\pm, %\label{eq73}
\end{gather*}
where $D^\pm$ are f\/irst order dif\/ferential operators of the form
\begin{gather*}
D^\pm=\frac{\hbar^{2}\beta^{2}}{2m}\coth \left(\frac{\beta x_-}{2}\right)\mp\frac{\beta \hbar^{2}}{m}\left(\partial_{x_-}+ \coth\left(\frac{\beta x_-}{2}\right) \partial_{x_+}\right),\qquad x_{\pm}=x\pm y,
\end{gather*}
we thus know that $\tilde H$ and $H$ are related by the intertwining relations:
\begin{gather}
{\tilde H} Q^+=Q^+ H, \qquad  H Q^-=Q^- \tilde H.\label{intert}
\end{gather}

It means, in particular, that if $\psi$ is an eigenstate of $H$, $Q^+ \psi$ is an eigenstate of $\tilde H$ with the same eigenvalue. In fact, it has been proven that the new Hamiltonian $\tilde H$ shares part of the energy spectrum of~$H$. More precisely, the only normalizable eigenstates of $\tilde H$ are given as (to simplify the developments, the set $(n_{x},n_{y})$  has been renamed $(n,m)$)
\begin{gather}
\tilde \Psi_{n,m}^A= Q^+ \Psi_{n,m}^{A}=(E_m- E_n) \Psi_{n,m}^{S}+D^+ \Psi_{n,m}^{A},\label{functionstilde}
\end{gather}
where
 \begin{gather*}
 \Psi_{n,m}^{S,A}=\frac{1}{\sqrt 2}(\psi_n^\nu (x)\psi_m^\nu (y)\pm \psi_m^\nu (x)\psi_n^\nu (y)),
 \end{gather*}
  with $\psi_n^\nu (x)$, $\psi_m^\nu (y)$ the well-known eigenfunctions of the 1D Morse system  \cite{Dong}. The functions~\eqref{functionstilde} are in fact antisymmetric in $(n,m)$.
The corresponding eigenvalues are given by $E_{n,m}=\eqref{energy12}$, for $n,m=0,1,2,\dots,[p]$ but now $n\neq m$ and $n\neq m\pm 1$ since $\tilde \Psi_{n,n}^A=0$ and $\tilde \Psi_{n,n+1}^A=0$.
 It has been shown~\cite{IN} that we thus get the complete spectrum for~$\tilde H$ that is a~subset of the spectrum of~$H$.

Let us summarize the properties involved in this context that will be useful for our interpretation of the arithmetical degeneracies. From~\eqref{intert}, we get  $[Q^- Q^+, H]=0$ and $Q^- Q^+ \Psi_{n,m}=r_{n,m} \Psi_{n,m}$ with
 \begin{gather}
r_{n,m}=(E_{m+1}- E_n)(E_{m-1}- E_n)=\frac{\hbar^{2}\epsilon^{2}}{4}\big((m-n)^2-1\big)\big((2p-m-n)^2-1\big),\label{rnm}
\end{gather}
where $r_{n,n+1}=r_{n+1,n}$.
In fact, $Q^- Q^+$ is a fourth order dif\/ferential operator acting on the eigenstates $\Psi_{n,m}$ of $H$ that could be written as
 \begin{gather}
Q^- Q^+=(H_{x}-H_{y})^2+2 H+I . \label{qmoinsplus}
\end{gather}
 Moreover, we get $[Q^+ Q^-, {\tilde H}]=0$ and $Q^+ Q^- \tilde \Psi_{n,m}^A=r_{n,m}\tilde \Psi_{n,m}^A$.  The operator $Q^+ Q^-$ is again a~fourth order dif\/ferential operator acting on the eigenstates $\tilde \Psi_{n,m}^A$ of~${\tilde H}$.

The interesting consequence is that we have found a superalgebra constructed from both Hamiltonians and the corresponding supercharges as follows. We introduce the generators
\begin{gather*}
{\cal Q}^+=
\left(\begin{array}{ccc}
0& Q^+\\
0&0
\end{array}\right),\qquad {\cal Q}^-=
\left(\begin{array}{ccc}
0& 0\\
Q^-&0
\end{array}\right), \qquad {\cal H}=\left(\begin{array}{ccc}
\tilde H& 0\\
0&H
\end{array}\right)%\label{eq79}
\end{gather*}
that satisfy
\begin{gather*}
[{\cal H},{\cal Q}^\pm]=0, \qquad \{{\cal Q}^+,{\cal Q}^-\}={\cal R},%\label{eq80}
\end{gather*}
where ${\cal R}$ commutes with ${\cal H}$. Thus the set $\{{\cal H},{\cal R},{\cal Q}^+, {\cal Q}^-\}$ closes a superalgebra. Note that ${\cal R}$ is a fourth order dif\/ferential operator which will be useful to explain the arithmetical degeneracies of the spectra of both $H$ and $\tilde H$.

We are now ready to proceed to the analysis of arithmetical degeneracies of the 2D Morse systems.
Let us consider, for example, the Hamiltonian $H$ and assume that it admits a double arithmetical degeneracy in the energy spectrum  for the couples $(n_1,m_1)$ and $(n_2,m_2)$. It means that  $E_{n_1,m_1}=E_{n_2,m_2}$ (avoiding the permutation symmetry and the identical relation). From the supersymmetric context, we have introduced a new constant of motion ${\cal R}$ with eigenvalues $r_{n,m}=\eqref{rnm}$. We are thus showing that  $r_{n_1,m_1}\neq r_{n_2,m_2}$.

Let us f\/irst set $p-n_i=k_i$ and $p-m_i=l_i$. Since $E_{n_1,m_1}=E_{n_2,m_2}$, we get
\begin{gather*}
k_1^2+l_1^2=k_2^2+l_2^2.%\label{eq81}
\end{gather*}
If we express $k_1^2$ in terms of the other quantities, we see that the dif\/ference $r_{n_1,m_1}- r_{n_2,m_2}$ becomes
\begin{gather*}
r_{n_1,m_1}- r_{n_2,m_2}= 4 \big(l_1^2-l_2^2\big)\big(l_1^2-k_2^2\big)\\
\hphantom{r_{n_1,m_1}- r_{n_2,m_2}}{} =4(m_2-m_1)(n_2-m_1)(2p-(m_2+m_1))(2p-(n_2+m_1)).%\label{eq82}
\end{gather*}
To get  $r_{n_1,m_1}- r_{n_2,m_2}=0$, we have 4 possibilities:
\begin{enumerate}\itemsep=0pt
\item[1)] $m_1=m_2$ and this leads to $E_{n_1,m_1}=E_{n_2,m_1}\iff n_1=n_2$ (identity, excluded);

\item[2)] $m_1=n_2$ and this leads to $E_{n_1,m_1}=E_{m_1,m_2}\iff n_1=m_2$ (symmetry permutation, excluded);

\item[3)] $m_1+m_2=2p$ and this leads to $m_1=m_2=p$ since $m_i\leq p$ and $E_{n_1,p}=E_{n_2,p}\iff n_1=n_2$ (identity, excluded);

\item[4)] $m_1+n_2=2p$ and this leads to $m_1=n_2=p$ and $E_{n_1,p}=E_{p,m_2} \iff n_1=m_2$ (symmetry permutation excluded).
\end{enumerate}

So we conclude that $r_{n_1,m_1}\neq r_{n_2,m_2}$ and the arithmetical degeneracies are explained by the existence of~${\cal R}$ since the preceding proof may be extended to multiple degeneracies.

\section{Conclusion}\label{section6}

In earlier works~\cite{Ma1, Ma2, FH}, we have presented some ways of getting integrals of motion and polynomial symmetry algebras in classical and quantum mechanics for multi-dimensional superintegrable systems. These quantities are obtained from 1D systems with polynomial ladder operators satisfying polynomial Heisenberg algebras. The integrals are products of ladder operators.

It was also observed in context of superintegrable systems that higher order ladder operators and thus higher order integrals of motion can be written in terms of f\/irst order and second order supercharges~\cite{Ma1, Ma3, Ma5}.
In particular, we constructed higher integrals of motion for an inf\/inite family of superintegrable systems and involving the f\/ifth Painlev\'e transcendent using products of second order supercharges~\cite{Ma5}. However, all these Hamiltonians were separable in cartesian coordinates. Let us mention that it has been shown very recently that a new non separable superintegrable system admits one third and one fourth order integrals of motion~\cite{PW}.

\looseness=1
In this paper, we have considered systems with non polynomial ladder operators. The method does not produce superintegrable systems and this is a limitation to the construction of~\cite{Ma1, Ma2,FH}. However, 2D integrable systems with non polynomial ladder operators can be constructed. In the classical case, we have obtained algebraically the trajectories and the condition for the existence of closed trajectories. In the quantum case, we have explained the degeneracies of the energy spectrum which involve, in particular, supersymmetric methods. We have applied these results for the classical and quantum inf\/inite well and Morse potentials.

Finally, let us observe that the new integral of motion $\cal R$, which is a fourth order dif\/ferential operator, is given as
\begin{gather*}
{\cal R}=\left(\begin{array}{ccc}
Q^+ Q^-& 0\\
0&Q^- Q^+\\
\end{array}\right),
\end{gather*}
and acts on the superspace of eigenfunctions $\{(0, \Psi_{n,n}^S)^T,\,  (\tilde \Psi_{n,m}^A,  \Psi_{n,m}^A)^T,\,  n> m; \, n,m=0,1,2$, $\dots,[p] \}$. We see that  $Q^- Q^+=\eqref{qmoinsplus}$ is quadratic in $H_x$ and $H_y$. The operator $Q^+ Q^-$ has a~similar expression but with additional terms. Indeed, we can show that
 \begin{gather*}
Q^+ Q^-=(H_{x}-H_{y})^2+2 \tilde H+I +[H_{x}-H_{y}, D^+ -D^-].
\end{gather*}
We could say that we have a nonlinear supersymmetry but, in fact, these last operators are not quadratic in $H$ or $\tilde H$.

We see that the supercharges $Q^\pm$  are related to $S=H_{x}-H_{y}$, the integral of motion obtained somewhat trivially  for Hamiltonians admitting the separation of variables in cartesian coordinates. The fourth order integral of motion $\cal R$ associated to the Morse Hamiltonian in 2D and related to the ``square'' of $Q$ is very interesting since one of the Hamiltonian ($\tilde H$) does not allow the separation of variables.

\subsection*{Acknowledgements}

The research of I.~Marquette was supported by a postdoctoral research fellowship from FQRNT of Quebec. V.~Hussin acknowledge the support of research grants from NSERC of Canada. Part of this work has been done while V.~Hussin visited Northumbria University (as visiting professor and sabbatical leave).

\newpage

\pdfbookmark[1]{References}{ref}
\LastPageEnding


\begin{thebibliography}{99}

\footnotesize\itemsep=0pt


\bibitem{Ma1}
Marquette I.,
Superintegrability and higher order polynomial algebras,
\href{http://dx.doi.org/10.1088/1751-8113/43/13/135203}{{\it J.~Phys.~A: Math. Theor.}} {\bf 43} (2010), 135203, 15~pages,
\href{http://arxiv.org/abs/0908.4399}{arXiv:0908.4399}.

\bibitem{Ma2}
Marquette I.,
Construction of classical superintegrable systems with higher order integrals of motion from ladder operators,
\href{http://dx.doi.org/10.1063/1.3448925}{{\it J.~Math. Phys.}} {\bf 51} (2010), 072903, 9~pages, 	
\href{http://arxiv.org/abs/1002.3118}{arXiv:1002.3118}.

\bibitem{FH}
Fern\'andez D.J., Hussin V.,
Higher-order SUSY, linearized nonlinear Heisenberg algebras and coherent states,
\href{http://dx.doi.org/10.1088/0305-4470/32/19/311}{{\it  J.~Phys.~A: Math. Gen.}} {\bf 32} (1999), 3603--3619.

\bibitem{CFNN}
Carbello J.M., Fern\' andez D.J., Negro J., Nieto L.M.,
Polynomial Heisenberg algebras,
\href{http://dx.doi.org/10.1088/0305-4470/37/43/022}{{\it  J.~Phys.~A: Math. Gen.}} {\bf 37} (2004), 10349--10362.

\bibitem{Ma3}
Marquette I.,
Supersymmetry as a method of obtaining new superintegrable systems with higher order integrals of motion,
\href{http://dx.doi.org/10.1063/1.3272003}{{\it J.~Math. Phys.}} {\bf 50} (2009), 122102, 10~pages,
\href{http://arxiv.org/abs/0908.1246}{arXiv:0908.1246}.

\bibitem{Junker}
Junker G.,
Supersymmetric methods in quantum and statistical physics,
{\it Texts and Monographs in Physics}, Springer-Verlag, Berlin, 1996.

\bibitem{Ma4}
Marquette I., Winternitz P.,
Superintegrable systems with third-order integrals of motion,
\href{http://dx.doi.org/10.1088/1751-8113/41/30/304031}{{\it  J.~Phys.~A: Math. Theor.}} {\bf 41} (2008), 304031, 10~pages,
\href{http://arxiv.org/abs/0711.4783}{arXiv:0711.4783}.

\bibitem{Delange}
De Lange O.L., Raab R.E.,
Operator methods in quantum mechanics, The Clarendon Press, Oxford University Press, New York, 1991.

\bibitem{DQ}
Delbecq C., Quesne C.,
Nonlinear deformations of su(2) and su(1,1) generalizing Witten's algebra,
\href{http://dx.doi.org/10.1088/0305-4470/26/4/001}{{\it  J.~Phys.~A: Math. Gen.}} {\bf 26} (1993), L127--L134.

\bibitem{EK1}
Eleonsky V.M., Korolev V.G.,
On the nonlinear generalization of the Fock method,
\href{http://dx.doi.org/10.1088/0305-4470/28/17/026}{{\it  J.~Phys.~A: Math. Gen.}} {\bf 28} (1995), 4973--4985.

\bibitem{EK2}
Eleonsky V.M., Korolev V.G.,
On the nonlinear Fock description of quantum systems with quadratic spectra,
\href{http://dx.doi.org/10.1088/0305-4470/29/10/004}{{\it  J.~Phys.~A: Math. Gen.}} {\bf 29} (1996), L241--L248.

\bibitem{GMK}
Ghosh A., Mitra P., Kundu A.,
Multidimensional isotropic and anisotropic $q$-oscillator models,
\href{http://dx.doi.org/10.1088/0305-4470/29/1/013}{{\it J.~Phys.~A: Math. Gen.}} {\bf 29} (1996), 115--124,
\href{http://arxiv.org/abs/hep-th/9511084}{hep-th/9511084}.

\bibitem{Quesne}
Quesne C.,
Comment: ``Application of nonlinear deformation algebra to a physical system with P\"oschl--Teller potential'' [{\it J. Phys.~A: Math. Gen.} {\bf 31} (1998), 6473--6481] by J.-L.~Chen, Y.~Liu and M.-L.~Ge,
\href{http://dx.doi.org/10.1088/0305-4470/32/38/401}{{\it  J.~Phys.~A: Math. Gen.}} {\bf 32} (1999), 6705--6710,
\href{http://arxiv.org/abs/math-ph/9911004}{math-ph/9911004}.

\bibitem{CZW}
Chen J.-L., Zhang H.-B., Wang X.-H., Jing H., Zhao X.-G.,
Raising and lowering operators for a two-dimensional hydrogen atom by an ansatz method,
\href{http://dx.doi.org/10.1023/A:1003661821241}{{\it Int. J. Theor. Phys.}} {\bf 39} (2000), 2043--2050.

\bibitem{CR}
Curado E.M.F., Rego-Monteiro M.A.,
Multi-parametric deformed Heisenberg algebras: a route to comple\-xi\-ty,
\href{http://dx.doi.org/10.1088/0305-4470/34/15/304}{{\it  J.~Phys.~A: Math. Gen.}} {\bf 34} (2001), 3253--3264,
\href{http://arxiv.org/abs/hep-th/0011126}{hep-th/0011126}.

\bibitem{DM}
Dong S.-H., Ma Z.-Q.,
The hidden symmetry for a quantum system with an inf\/initely deep square-well potential,
\href{http://dx.doi.org/10.1119/1.1456073}{{\it Amer. J. Phys.}} {\bf 70}  (2002), 520--521.

\bibitem{DK}
Daoud M., Kibler M.R.,
Fractional supersymmetry and hierarchy of shape invariant potentials,
\href{http://dx.doi.org/10.1063/1.2401711}{{\it J.~Math. Phys.}} {\bf 47} (2006), 122108, 11~pages,
\href{http://arxiv.org/abs/quant-ph/0609017}{quant-ph/0609017}.

\bibitem{Dong}
Dong S.-H.,
Factorization method in quantum mechanics,
{\it Fundamental Theories of Physics}, Vol.~150, Springer, Dordrecht, 2007.

\bibitem{CH}
Curado E.M.F., Hassouni Y., Rego-Monteiro M.A., Rodrigues L.M.C.S.,
Generalized Heisenberg algebra and algebraic method: the example of an inf\/inite square-well potential,
\href{http://dx.doi.org/10.1016/j.physleta.2008.01.086}{{\it Phys. Lett.~A}} {\bf 372} (2008), 3350--3355.

\bibitem{WYB}
Wang H.-B.,  Liu Y.-B.,
Realizing the underlying quantum dynamical algebra $SU(2)$ in Morse potential,
\href{http://dx.doi.org/10.1088/0256-307X/27/2/020301}{{\it Chinese Phys. Lett.}} {\bf 27} (2010), 020301, 4~pages.

\bibitem{IN}
Iof\/fe M.V., Nishnianidze D.N.,
Exact solvability of two-dimensional real singular Morse potential,
\href{http://dx.doi.org/10.1103/PhysRevA.76.052114}{{\it Phys. Rev.~A}} {\bf 76} (2007), 052114, 5~pages,
\href{http://arxiv.org/abs/0709.2960}{arXiv:0709.2960}.

\bibitem{KN}
Kuru S., Negro J.,
Factorizations of one dimensional classical systems,
\href{http://dx.doi.org/10.1016/j.aop.2007.10.004}{{\it Ann. Physics}} {\bf 323} (2008), 413--431,
\href{http://arxiv.org/abs/0709.4649}{arXiv:0709.4649}.

\bibitem{CKN}
Cruz y Cruz S., Kuru S., Negro J.,
Classical motion and coherent states for  P\"oschl--Teller  potentials,
\href{http://dx.doi.org/10.1016/j.physleta.2007.10.010}{{\it Phys. Lett.~A}} {\bf 372} (2008), 1391--1405.

\bibitem{DH}
Dello Sbarba L., Hussin V.,
Degenerate discrete energy spectra and associated coherent states,
\href{http://dx.doi.org/10.1063/1.2435596}{{\it J.~Math. Phys.}} {\bf 48} (2007), 012110, 15~pages.

\bibitem{AH1}
Angelova M.,  Hussin V.,
Generalized and Gaussian coherent states for the Morse potential,
\href{http://dx.doi.org/10.1088/1751-8113/41/30/304016}{{\it J.~Phys.~A: Math. Theor.}} {\bf 41} (2008), 30416, 13~pages.

\bibitem{AH2}
Angelova M., Hussin V.,
Squeezed coherent states and the Morse quantum system,
\href{http://arxiv.org/abs/1010.3277}{arXiv:1010.3277}.

\bibitem{DLF}
Dong S.H., Lemus R., Frank A.,
Ladder operators for the Morse potential,
\href{http://dx.doi.org/10.1002/qua.10038}{{\it Int. J. Quant. Chem.}} {\bf 86} (2002), 433--439.

\bibitem{BMQ}
Bagchi B., Mallik S., Quesne C.,
Inf\/inite square well and periodic trajectories in classical mechanics,
\href{http://arxiv.org/abs/physics/0207096}{physics/0207096}.

\bibitem{Slater}
Slater N.B.,
Classical motion under a Morse potential,
\href{http://dx.doi.org/10.1038/1801352a0}{{\it Nature}} {\bf 180} (1957), 1352--1353.

\bibitem{Itzykson}
Itzykson C., Luck J.M.,
Arithmetical degeneracies in simple quantum systems,
\href{http://dx.doi.org/10.1088/0305-4470/19/2/017}{{\it J.~Phys.~A: Math. Gen.}} {\bf 19} (1986), 211--239.

\bibitem{Ma5}
Marquette I.,
An inf\/inite family of superintegrable systems from higher order ladder operators and supersymmetry,
in Group 28: Physical and Mathematical Aspects of Symmetry: Proceedings of the 28th International Colloquium on Group-Theoretical Methods in Physics,
{\it J. Phys. Conf. Ser.}, to appear,
\href{http://arxiv.org/abs/1008.3073}{arXiv:1008.3073}.

\bibitem{PW}
Post S., Winternitz P.,
A nonseparable quantum superintegrable system in 2D real Euclidean space,
\href{http://arxiv.org/abs/1101.5405}{arXiv:1101.5405}.

\end{thebibliography}
\end{document}